\documentclass[aps,pra,twocolumn, superscriptaddress]{revtex4-2}
\usepackage{amsmath}
\usepackage{algorithm}
\usepackage{amsfonts}
\usepackage{graphicx}
\usepackage{algpseudocode}
\usepackage{xcolor}
\usepackage{mathtools}
\usepackage{tikz}
\usepackage{float}
\usepackage{adjustbox}
\usepackage{hyperref}

\graphicspath{{./figures}}

\makeatletter
\renewcommand{\ALG@name}{Protocol}
\makeatother

\makeatletter
\newcommand*{\centerfloat}{%
  \parindent \z@
  \leftskip \z@ \@plus 1fil \@minus \textwidth
  \rightskip\leftskip
  \parfillskip \z@skip}
\makeatother

\begin{document}

\raggedbottom

\title{Enhancing Quantum Key Distribution with Entanglement Distillation and Classical Advantage Distillation}

\author{Shin Sun}
\affiliation{Networked Quantum Devices Unit, Okinawa Institute of Science and Technology Graduate University, Okinawa, Japan}

\author{Kenneth Goodenough}
\affiliation{College of Information and Computer Science, University of Massachusetts Amherst, 140 Governors Dr, Amherst, Massachusetts 01002, USA}
\author{Daniel Bhatti}
\affiliation{Networked Quantum Devices Unit, Okinawa Institute of Science and Technology Graduate University, Okinawa, Japan}
\author{David Elkouss}
\affiliation{Networked Quantum Devices Unit, Okinawa Institute of Science and Technology Graduate University, Okinawa, Japan}

\begin{abstract}
Realizing secure communication between distant parties is one of quantum technology’s main goals.
Although quantum key distribution promises information-theoretic security for sharing a secret key, the key rate heavily depends on the level of noise in the quantum channel.
To overcome the noise, both quantum and classical techniques exist, i.e., entanglement distillation and classical advantage distillation.
So far, these techniques have only been used separately from each other.
Herein, we present a two-stage distillation scheme concatenating entanglement distillation with classical advantage distillation.
For advantage distillation, we utilize a fixed protocol, specifically, the \textit{repetition code}; in the case of entanglement distillation, we employ an enumeration algorithm to find the optimal protocol. 
We test our scheme for different noisy entangled states and demonstrate its quantitative advantage: our two-stage distillation scheme achieves finite key rates even in the high-noise regime where entanglement distillation or advantage distillation alone cannot afford key sharing. We also calculate the security bounds for relevant QKD protocols with our key distillation scheme and show that they exceed the previous security bounds with only advantage distillation. Since the advantage distillation part does not introduce further requirements on quantum resources, the proposed scheme is well-suited for near-term quantum key distribution tasks. 

\end{abstract}

\maketitle

\section{Introduction}
Quantum technologies open the possibility of implementing communication tasks with qualitative advantages with respect to the current communication networks. Notably, they enable the distribution of secure keys without making computational assumptions \cite{bennett2014quantum,ekert1991quantum}. The security of keys shared through quantum key distribution (QKD) protocols does not rely on the assumption of the adversary's computational capability but is solely based on the validity of quantum mechanics. 

Typically, the security of QKD protocols can be related to the presence of entanglement \cite{ekert1991quantum}. The intuition is that if both parties share (pure) maximally entangled states, they can then obtain correlated outcomes that are independent of any third party, which renders any eavesdropping attack impossible. However, in the presence of realistic devices, quantum information processing is necessarily noisy. This noise degrades the rate at which a quantum key distribution protocol produces secure keys~\cite{xu2020secure}.

There are several possible strategies to overcome noise. If the key is produced by measuring an entangled state, entanglement distillation (ED) \cite{bennett1996mixed,deutsch1996quantum} can probabilistically transform a number of noisy entangled states into a smaller number of more entangled states. Entanglement distillation has already been demonstrated in experiments, e.g., \cite{kwiat2001experimental,pan2003experimental,reichle2006experimental,kalb2017entanglement, hu2021long}. However, state-of-the-art quantum communication hardware can only store a handful of states, limiting the amount of entanglement distillation possible. 

Alternatively, it is possible to act directly on the classical outcomes obtained by measuring the quantum resources. Advantage distillation (AD)~\cite{maurer1993secret,christandl2007unifying} is a purely classical procedure that transforms a bit string probabilistically on which an eavesdropper has a large amount of information into a smaller bit string on which the eavesdropper has less information.
While there is no limitation on the amount of advantage distillation that the legitimate key distribution parties can apply, it is a less efficient process than entanglement distillation with fundamental differences in the amount of noise that it can overcome \cite{gottesman2003proof, bae2007key}.

Herein, we propose a state-aware two-stage key distillation scheme by concatenating entanglement distillation with advantage distillation. An obvious advantage of this scheme is that it does not require additional quantum resources than those required to perform the entanglement distillation in the first stage alone.
Our scheme builds on a recent enumeration of entanglement distillation algorithms \cite{jansen2022enumerating, goodenough2024near}, which allows us to optimize the two-stage scheme for a given starting state. 
In particular, by recognizing the equivalence of an advantage distillation protocol and a CNOT-based entanglement distillation protocol, we generalize the enumeration scheme to concatenated entanglement distillation with advantage distillation (ED+AD) protocols. 
We quantitatively demonstrate that this two-stage distillation scheme can achieve non-zero key rates at a heavy noise regime, whereas using either stage alone could not achieve positive key rates.

The paper is structured as follows: in Sec.~\ref{sec:entangled_QKD}, we review the basics of entanglement-based quantum key distribution. In Sec.~\ref{sec:clifford_protocols}, we introduce the mathematical structure of bi-local Clifford distillation protocols, which form an important subset of entanglement distillation protocols; further mathematical details will be provided in the appendix. In Sec.~\ref{sec:advantage_distillation}, we review the idea of classical advantage distillation and discuss its connection to bi-local Clifford distillation protocols. In Sec.~\ref{sec:ED_AD}, we describe the concatenated ED+AD protocols and the method we used to enumerate all possible such protocols for the optimization of QKD performance. In the results section, we present and discuss the quantitative performance of the ED+AD protocols for two particular noisy entangled states. Importantly, we derive the critical quantum bit error rates (QBER) for two standard QKD protocols with the ED+AD key distillation scheme. We demonstrate the ED+AD strategy strictly tolerates more noise than using advantage distillation alone.

\section{Background}
Here, we review the technical background of entanglement-based key distribution and two main techniques employed in this paper, namely bi-local Clifford entanglement distillation and classical advantage distillation.
\subsection{Entanglement-based quantum key distribution} \label{sec:entangled_QKD}
Quantum key distribution protocols can be categorized into the prepare-and-measure schemes (e.g., the original form of the BB84 protocol~\cite{bennett2014quantum} and the six-state protocol~\cite{bruss1998optimal}) or the entanglement-based schemes (e.g., the E91 protocol~\cite{ekert1991quantum}). 
However, it has been shown that these two categories are mathematically equivalent~\cite{shor2000simple}. For example, the prepare-and-measure BB84 protocol can be performed in the following way based on pre-shared entanglement. 

\begin{algorithm}[H]
  \caption{The entanglement-based BB84 protocol.}\label{alg:entanglement_based_BB84}
   \begin{algorithmic}[1]
   \State Alice and Bob share entangled pairs in the form of $|\Phi^+\rangle = \frac{1}{\sqrt{2}} \left(|00\rangle + |11\rangle \right)$.
   \State Both parties measure each shared entangled qubit in one of the two randomly chosen bases, $\mathcal{B}_1 = \{ |0\rangle, |1\rangle \}$ or $\mathcal{B}_2 = \{ \frac{1}{\sqrt{2}} \left( |0\rangle + |1\rangle \right), \frac{1}{\sqrt{2}} \left( |0\rangle - |1\rangle \right) \}$.
   \State Both parties announce the basis they used for the measurement. The measurement outcomes corresponding to the same basis are kept.
   \State A portion of the raw key string is announced and exchanged to assess the error rates.
   \State Depending on the observed error rates, both parties decide if the unannounced part of the key is kept.
   \State Information reconciliation and privacy amplification are performed to obtain secure keys.~\cite{cachin1997linking}
   \end{algorithmic}
\end{algorithm}

In this entanglement-based picture, any eavesdropping attack attempts can be modeled as noise in the shared entanglement. This makes the security analysis more convenient compared to the prepare-and-measure picture~\cite{scarani2009security}.

\subsection{Bi-local Clifford entanglement distillation protocols} \label{sec:clifford_protocols}
\begin{figure}[htbp]
\centerfloat
\includegraphics[width=\linewidth]{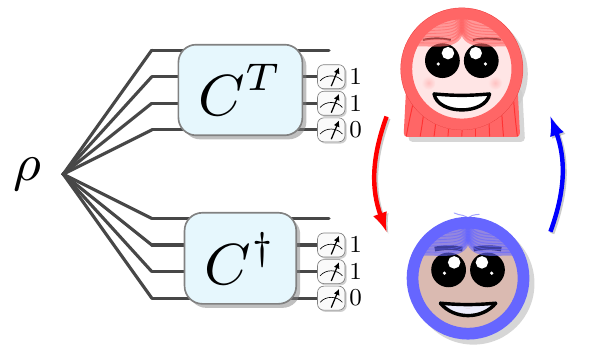}
  \caption{\textbf{Illustration of a bi-local Clifford entanglement distillation protocol.} Alice and Bob first share some noisy entangled pairs. They then perform local Clifford operations ($C^T$ and $C^\dagger$) on their share of qubits. They measure all pairs except the first pair. If the measurement outcomes of the corresponding entangled pairs coincide, they keep the first unmeasured pair. Otherwise, they discard it. }
  \label{fig:clifford}
\end{figure}

Entanglement distillation refers to turning a number $m$ of noisy entangled pairs into a smaller number $k$ of more entangled pairs by means of local quantum operations and classical communication~\cite{bennett1996mixed}. Due to their mathematical structure, the bi-local Clifford entanglement distillation protocols have been extensively studied~\cite{jansen2022enumerating, goodenough2024near, zang2024no}. Here, we restrict our attention to bi-local Clifford protocols taking $m$ pairs to 1 pair ($m$-1 protocol); see Fig.~\ref{fig:clifford} for an illustration of a $4$-$1$ bi-local Clifford entanglement distillation protocol. More precisely, an $m$-$1$ bi-local Clifford entanglement distillation protocol is defined as follows:
\begin{algorithm}[H]
  \caption{A $m$-$1$ bi-local Clifford entanglement distillation protocol.}\label{alg:bi-local_clifford}
   \begin{algorithmic}[1]
   \State Alice and Bob first share $m$ entangled pairs in the form of $|\Phi^+\rangle = \frac{1}{\sqrt{2}} \left(|00\rangle + |11\rangle \right)$ through noisy quantum channels.
   \State Alice performs a Clifford operation $C^T$ on her $m$ qubits, and Bob performs a corresponding $C^\dagger$ Clifford operation on his $m$ qubits.
   \State They perform measurements on all qubits except the first qubit.
   \State They publicly announce the measurement outcomes and keep the first qubit if the measurement outcomes of corresponding pairs are coincident.
   \end{algorithmic}
\end{algorithm}

There are many well-known protocols that belong to the bi-local Clifford class. For instance, the DEJMPS protocol (also known as Oxford protocol)~\cite{deutsch1996quantum} is an example of a $2$-$1$ bi-local Clifford protocol. In DEJMPS, the Clifford operation is given by $C^T =\left(\text{CNOT}_{12}\right) \left(R_{x,1}(\frac{\pi}{2}) \otimes R_{x,2}(\frac{\pi}{2})\right)$; where the $\text{CNOT}_{12}$ is performed on local qubit pairs and $R_{x,i}$ is an x-axis rotation on the $i$-th qubit.

Bi-local Clifford protocols are a very general class of protocols. Optimizing over this class can be done by enumerating all elements of the Clifford group. However, the number of Clifford gates grows super exponentially with the number of qubits~\cite{jansen2022enumerating}. Fortunately, it is possible to simplify the enumeration by building on the mathematical correspondence between Clifford operations acting on the $n$-qubit Pauli group (ignoring the phase information) and the symplectic group $\textrm{Sp}(2n, \mathbb{Z}_2)$ with its natural group action (readers are referred to appendix~\ref{sec:appendix_binary_picture} for the details). In particular,  Clifford gates $C_1$ and $C_2$, when used to perform the entanglement distillation, could lead to the same distillation statistics (i.e., output fidelity and success probability)~\cite{jansen2022enumerating,singal2023counting}. This observation enables enumerating all different bi-local Clifford protocols by enumerating an element in each distillation equivalence class. %
This reduction allows an enumeration algorithm to find the optimal protocol for a small number of shared noisy entangled pairs.

\subsection{Advantage distillation and its relation to entanglement distillation} \label{sec:advantage_distillation}
Advantage distillation (AD) is a well-known technique in secret key distillation scenarios. It allows Alice and Bob to enhance the correlation between their shared random variable and reduce its correlation to a third party~\cite{maurer1993secret}. It is designed so that Alice and Bob can gain an ``advantage'' over the eavesdropper---Eve---by recognizing and keeping the part of the shared random variables that has a higher correlation between Alice and Bob than Eve. Advantage distillation typically uses parity calculation and two-way communication to decide if a certain part of the key is kept or not~\cite{bae2007key,liu2003practical}. 

We now provide an example of an AD protocol. Let $A^N$ and $B^N$ be $N$-bit correlated bit strings belonging to Alice and Bob, respectively. Alice generates a random bit $s$ and adds it bitwise to a block of $m\leq N$ bits ($a^m$) to obtain $C$ (which contains parity information of the block and the random bit). She then publicly sends $C$ to Bob via an authenticated channel. Bob calculates the bitwise addition of $C$ and his corresponding block ($b^m$). If the results are all-$0$ or all-$1$, Bob sends an accept signal back to Alice, and a correlated bit is obtained. The process is repeated several rounds until a key string with a suitable size is obtained. 

The action of calculating the parity can be phrased as applying CNOT gates (which are Clifford gates) to relevant qubits. This representation and the requirement of correlated results enable us to cast an advantage distillation protocol to a bi-local Clifford entanglement distillation protocol only comprised of CNOT gates. Thus, the advantage distillation in the form of a quantum circuit could also be viewed as applying a classical linear error detection code in the computational basis.

In this paper, we only consider $n$ to $1$ advantage distillation protocols, which take $n$ entangled pairs, calculate the parities locally, and announce $n-1$ pairs to decide if the remaining one is kept or not.

\section{ED+AD key distillation protocols}\label{sec:ED_AD}
For entanglement-based quantum key distribution, both entanglement distillation and advantage distillation increase the correlation of the shared key string. However, the former requires coherent quantum operations on the entangled states, while the latter can be performed after the quantum measurement and with the help of classical communication. Intuitively, entanglement distillation has the potential to increase the output fidelity with respect to a particular Bell state ($|\Phi^+\rangle$, in this paper) by correcting both bit-flip and phase-flip errors. On the other hand, the advantage distillation in the computational basis aims to correct the bit-flip error only. 

Herein, we describe the structure of an $m$-$n$-$1$ ED+AD concatenated protocol for entanglement-based quantum key distribution in Protocol~\ref{alg:ed_ad}.

\begin{figure}[htbp]
\centerfloat
\includegraphics[width=\linewidth]{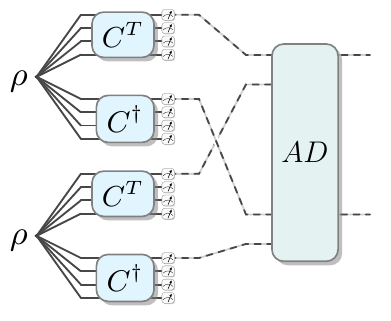}
  \caption{\textbf{Illustration of the ED+AD protocol.} The noisy entangled pairs shared by Alice and Bob first undergo entanglement distillation. The successfully distilled pairs are measured. The measurement outcomes (dashed lines) become the input to the advantage distillation protocol. The illustration corresponds to a 4-2-1 protocol, e.g., a concatenation of a 4-1 entanglement distillation protocol with a 2-1 advantage distillation protocol.}
  \label{fig:ed_ad}

\end{figure}

Note that the entanglement distillation stages, as shown in Fig.~\ref{fig:ed_ad}, do not have to be successful at the same time. Since the advantage distillation is performed offline based on the measurement outcome, the individual entanglement distillation steps can be executed in a repeat-until-success manner.

\begin{algorithm}[H]
    \caption{$m$-$n$-$1$ ED+AD key distillation protocol.}\label{alg:ed_ad}
     \begin{algorithmic}[1]
     \State Alice and Bob try to share at least $mn$ entangled qubit pairs in state $|\Phi^+\rangle = \frac{1}{\sqrt{2}} \left( |00\rangle + |11\rangle \right)$.
     
     \State Alice and Bob perform $m$-$1$ Clifford entanglement distillation protocols at least $n$ times to obtain $n$ entangled pairs with higher fidelity with respect to $|\Phi^+\rangle$.

     \State Both parties perform measurement on the $n$ pairs in the computational basis.

     \State Both parties perform $n$-$1$ advantage distillation protocol to decide if the remaining bit is kept or not.
     \end{algorithmic}
\end{algorithm}

\begin{table}[]

  \begin{tabular}{|c|c|c|}
  \hline
  Protocol name & ED  & AD  \\ \hline
  2-1 & 2-1 & \emph{none} \\ \hline
  3-1 & 3-1 & \emph{none} \\ \hline
  4-1 & 4-1 & \emph{none} \\ \hline
  2-2-1 & 2-1 & 2-1 \\ \hline
  2-3-1 & 2-1 & 3-1 \\ \hline
  3-2-1 & 3-1 & 2-1 \\ \hline
  3-3-1 & 3-1 & 3-1 \\ \hline
  2-4-1 & 2-1 & 4-1 \\ \hline
  4-2-1 & 4-1 & 2-1 \\ \hline
  \end{tabular}
  \caption{\label{tab:table1}\textbf{Distillation protocols and their constituents.} In this paper, we studied nine different distillation protocols. Their constituent protocols are summarized in this table.}
  \end{table}

\begin{figure*}[htbp]
  \centering
  \includegraphics[width=.9\linewidth]{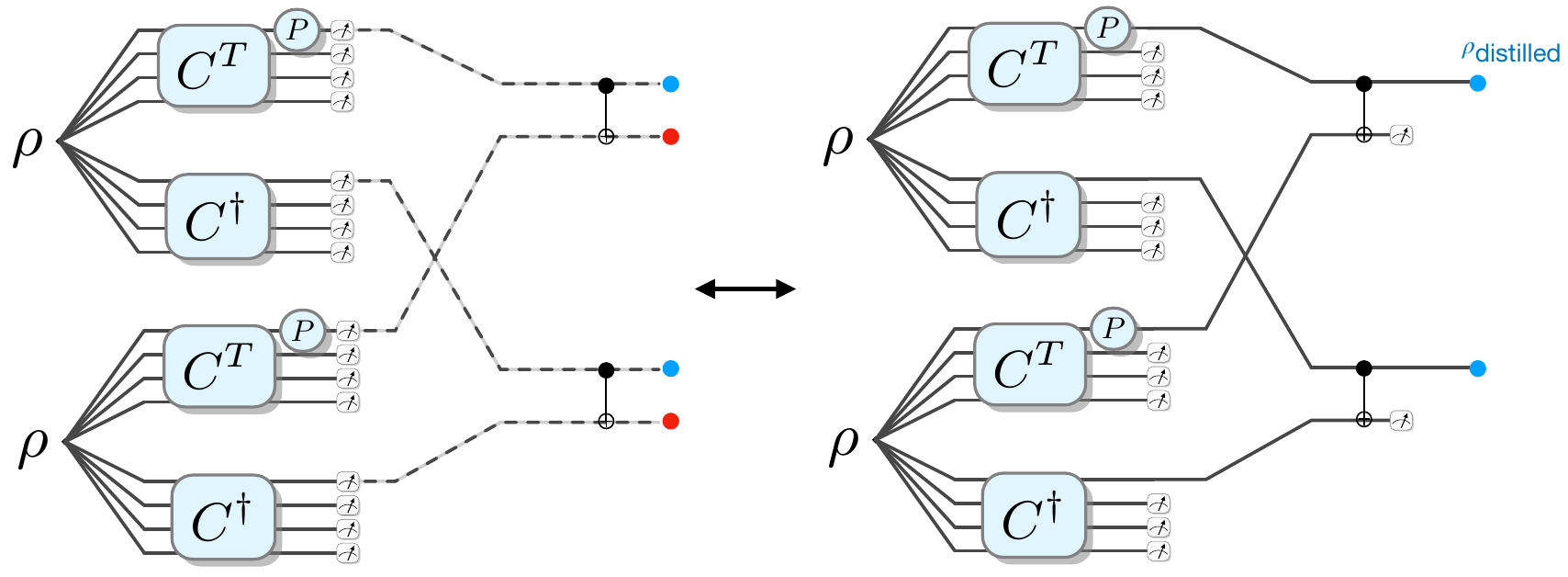}
  \caption{\textbf{Equivalence of ED+AD and a larger entanglement distillation protocol.} The left procedure is the QKD protocol based on two-stage distillation. $P$ is a local Pauli operator chosen from $\{I,X,Y,Z \}$. The red bits are compared via an authenticated channel. If the red bits have the same values, the blue bits are kept as raw secret bits. Since the CNOTs in computational basis commute with the measurement, we use the equivalent quantum picture on the right side to calculate the key rates with the distilled output $\rho_\text{distilled}$. The protocol shown here is a $4$-$2$-$1$ protocol defined in the paper.}
  \label{fig:equivalence_edad}

\end{figure*}

One key point for concatenating two distillation protocols is that the advantage distillation is performed after the quantum measurement. This means that the ED+AD protocol does not require additional quantum resources than those required for performing entanglement distillation alone. However, we note the following equivalence between advantage distillation protocols and a CNOT-only Clifford distillation protocol. As shown in Fig.~\ref{fig:equivalence_edad}, since the CNOTs performed in the computational basis commute with measurements, we can postpone the measurements to the end and use the quantum state before the measurement to calculate relevant figures of merit for QKD protocols. This equivalence also allows us to view an ED+AD protocol as an $mn$-$1$ distillation without having to use an $mn$-qubit quantum computer. 

In this paper, since we only consider the $n$-$1$ advantage distillation, the only non-trivial $n$-bit to $1$-bit error detection code is the repetition code~\cite{wootton2018repetition}. For example, the advantage distillation in Fig.~\ref{fig:equivalence_edad} corresponds to a $2$-$1$ repetition code. Therefore, we do not require an enumeration of the AD part of the protocol in order to optimize over $m$-$n$-$1$ ED+AD protocols.

We employ the following enumeration procedure to find the optimal protocol for a given noisy entangled state. A transversal $\mathcal{C}_m$ is a set that contains one bi-local distillation protocol from each distillation equivalence class (see appendix~\ref{sec:appendix_binary_picture}).
  \begin{algorithm}[H]
    \caption{Enumerating $m$-$n$-$1$ ED+AD protocols.}\label{alg:enumeration}
     \begin{algorithmic}[1]
     \State Generate transversal $\mathcal{C}_m$ of the $m$-$1$ Clifford distillation protocols~\cite{jansen2022enumerating}.
     
     \State Generate the advantage distillation protocol $A_n$ that corresponds to the repetition code $[n,1]$.

     \State Given a noisy state $\rho_\text{input}$, perform the following enumeration.
     \For{Clifford protocol $C_m$ in $\mathcal{C}_m$}
       \For {Single-qubit Pauli operator $P$}

 Calculate the distilled state $\rho_\text{distilled}$ by concatenating $C_m$, $P$, and $A_n$.
  
      \EndFor
        \EndFor

     \end{algorithmic}
\end{algorithm} 
For the purpose of entanglement distillation, 
if two protocols produce a distilled state up to the same distillation statistics, the two protocols are considered to be equivalent (see appendix~\ref{sec:appendix_binary_picture}). However, in our concatenation, we append one more distillation protocol to the output from the first Clifford distillation protocol. In this case, two distilled states with the same distillation statistics could lead to different final distilled states after the second stage, i.e., advantage distillation. Therefore, to consider the most general form of concatenating two protocols, we include the possibility of performing a local Pauli operation between the action of two distillation protocols. In the real implementation of the Clifford distillation protocol, the Pauli permutation part can be absorbed into the previous Clifford gate.

\section{Results}
In this section, we analyze the quantitative performance of the ED+AD $m$-$n$-1 protocols for two standard noise models. We enumerated all possible ED protocols in combination with the repetition-code AD protocol to find the optimal ED+AD protocol for the two specific noisy states that result from applying the corresponding noise model. In appendix~\ref{sec:appendix_comparison_DEJMPS}, we also show that the optimal protocol found by the enumeration performs better than naively using the DEJMPS protocol in the entanglement distillation stage. Finally, we discuss the security bounds for the ED+AD key distillation scheme and show that they outperform previous security bounds where only advantage distillation is considered.

\subsection{Depolarizing noise}
We first analyze the performance of the ED+AD protocols in the case of the Werner input state (Eqn.~\ref{eqn:werner})~\cite{dur2005standard}, which results from applying depolarizing noise on the Bell state $|\Phi^+\rangle$. The definition of the Bell states is given in Eqn.~\ref{eqn:bell_states}. The Werner state is a highly symmetric quantum state that is parametrized by a single parameter (i.e., the input fidelity $F$). The depolarizing noise can be seen as a worst-case noise model from a standard twirling argument~\cite{bennett1996mixed}.

\begin{align}\label{eqn:werner}
  \rho_\text{Werner} = F |\Phi^+\rangle \langle \Phi^+ | &+ \frac{1-F}{3} \left( |\Psi^+\rangle \langle \Psi^+ | + |\Psi^-\rangle \langle \Psi^- |\right. \nonumber \\ &+\left. |\Phi^-\rangle \langle \Phi^- | \right).
\end{align}

\subsubsection{Achievable fidelity}\label{sec:depolarizing_fidelity}

\begin{figure*}[htbp]
  \centering
  \includegraphics[width=.9\linewidth]{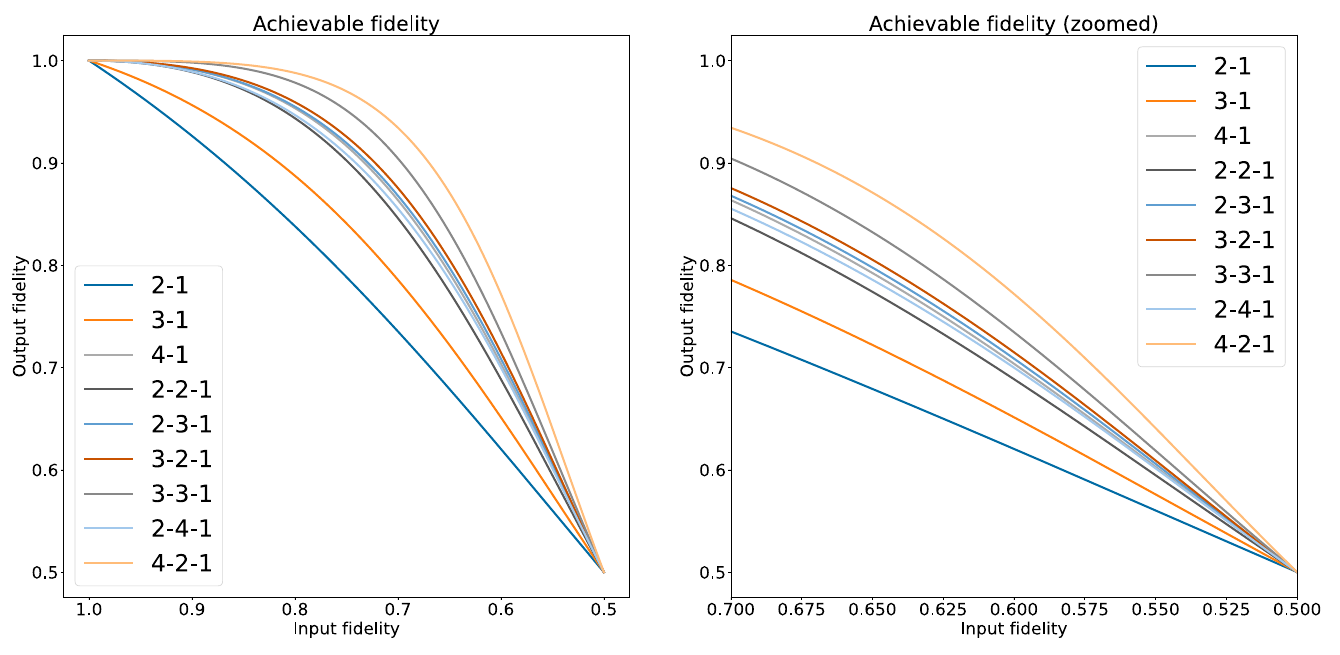}
  \caption{\textbf{Achievable fidelity - depolarizing noise.} For each input fidelity, we enumerate all possible distillation protocols and plot the optimal fidelity achieved by the distillation protocol. The $m$-$1$ protocols are the ED-only protocols. The $m$-$n$-$1$ protocols are ED+AD protocols.}
  \label{fig:depolarizing_fid}
\end{figure*}

We first calculate the achievable fidelity of the state $\rho_\text{distilled}$ ($F= \langle \Phi^+| \rho_\text{distilled} | \Phi^+\rangle $) for nine different distillation protocols in Fig.~\ref{fig:depolarizing_fid}. The $m$-$1$ protocols are the ED-only protocols. The $m$-$n$-$1$ protocols are ED+AD protocols.

The $m$-$n$-1 ED+AD protocols consistently perform better than their $m$-1 ED-only counterparts. It is interesting that the protocol $3$-$2$-$1$ achieves a higher fidelity than the $2$-$3$-$1$ protocol. This implies that using the more flexible $3$-$1$ entanglement distillation protocols in the first ED stage is more advantageous in terms of producing more entangled distilled states. Unexpectedly, the $2$-$4$-$1$ protocol achieves worse fidelity than the $2$-$3$-$1$ and $2$-$2$-$1$ protocols. This rather counter-intuitive behavior has its origin in the fact that, unlike entanglement distillation, advantage distillation only corrects for bit-flip errors in the computational basis. Therefore, the phase error component $|\Phi^-\rangle$ can be amplified during the distillation. This phenomenon is discussed in more detail in appendix~\ref{sec:appendix_repetition_code}, where we provide analytical expressions of performing AD-only. Note that for a fidelity of $\frac{1}{2}$ or less, the Werner state is separable, and no entanglement can be distilled.

\subsubsection{BB84 key rates}\label{sec:depolarizing_keyrates}
To consider the performance of the ED+AD scheme for key distribution, we calculate the asymptotic BB84 key rates~\cite{scarani2009security,lo2005efficient} for different protocols as follows.
\begin{align}
 r_\text{BB84} = \max \left\lbrace \frac{p_\text{ED} p_\text{AD}}{mn} \left( 1-H(p_\text{bit}) - H(p_\text{phase}) \right),0 \right\rbrace,
\end{align}
where $p_\text{ED}$ ($p_\text{AD}$) is the success probability for the entanglement distillation protocol (the success probability for the advantage distillation protocol), $p_\text{bit} (p_\text{phase})$ the probability of the bit-flip (phase-flip) error for the distilled state, and $H(\cdot)$ is the usual binary entropy function. If the ED-only protocol is used, $p_\text{AD}$ and $n$ are set to 1.

\begin{figure*}[htbp]
  \centering
  \includegraphics[width=.9\linewidth]{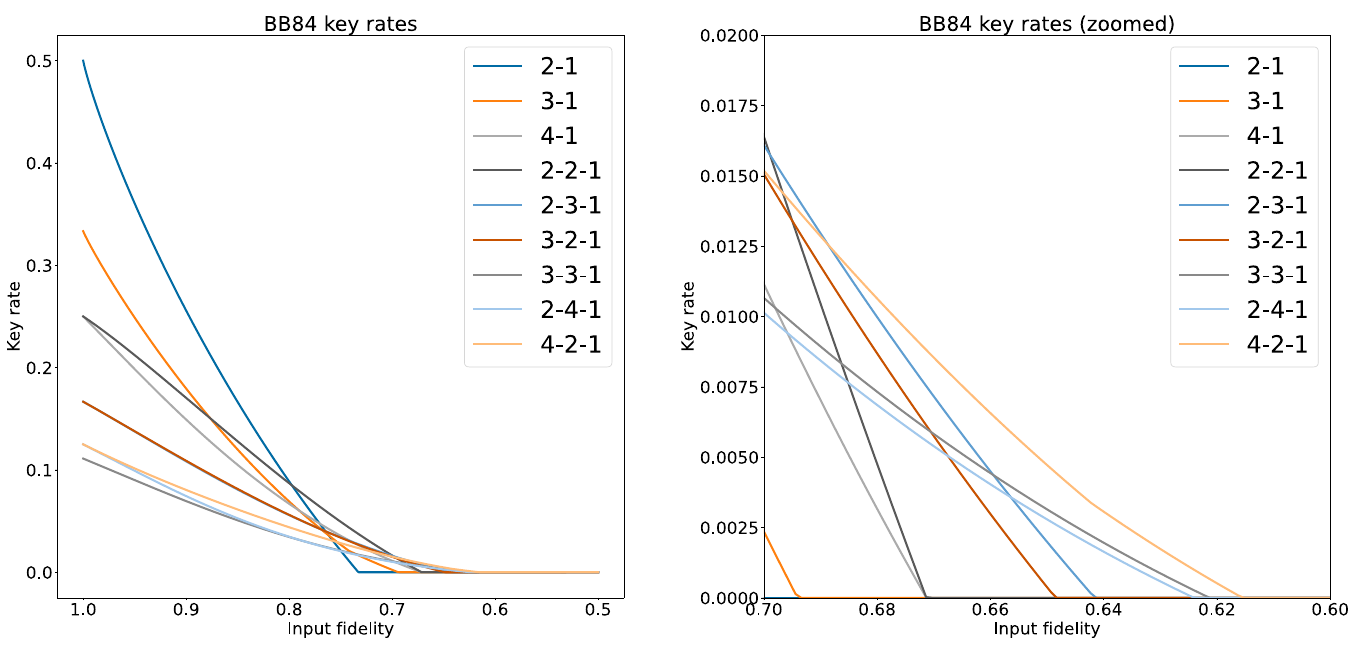}
  \caption{\textbf{Achievable BB84 asymptotic key rates - depolarizing noise.} For each input fidelity, we enumerate all possible distillation protocols and plot the optimal key rate achieved by the distillation protocol.}
  \label{fig:depolarizing_keyrates}

\end{figure*}

In the low noise regime, the ED-only protocols and $m$-$n$-$1$ protocols with smaller $m$ and $n$ perform better. This is due to the overhead incurred by the distillation. However, in the high noise regime, the advantage of ED+AD protocols begins to emerge. In the zoomed-in version of Fig.~\ref{fig:depolarizing_keyrates}, if the input fidelity is lower than around $0.67$, the ED-only protocols cannot achieve finite key rates, whereas several ED+AD protocols can still achieve finite key rates.

Rather surprisingly, in this noise regime, the protocol $2$-$3$-$1$ performs better than protocol $3$-$2$-$1$. This implies that in the case of depolarizing noise and using a quantum computer that can process at most $3$ qubits as input, using a smaller entanglement distillation ($m=2$) more often ($n=3$) is more advantageous than using a larger entanglement distillation protocol ($m=3$) and performing advantage distillation fewer times ($n=2$). The reason for the rather counter-intuitive situation is as follows: unlike the achievable fidelity calculation, when calculating the achievable key rates, the success probability of the distillation is also important. Although the $3$-$2$-$1$ protocols can, in principle, distill a more entangled state than $2$-$3$-$1$ protocols, the lower overall success probability makes the former less advantageous for performing QKD. In fact, this poses good news for experimental implementation since the larger entanglement distillation protocols require more qubits and longer coherence times in the quantum computer. %

\subsection{Pure dephasing noise}
As a second case, we consider pure dephasing noise (Eqn.~\ref{eqn:dephase})~\cite{clerk2010introduction}. This kind of noise arises naturally in various physical implementations of generating entanglement. 

\begin{align}\label{eqn:dephase}
  \rho_\text{dephase} = (1-p_\text{phase}) |\Phi^+\rangle \langle \Phi^+ | + p_\text{phase}   |\Phi^-\rangle \langle \Phi^- | \ .
\end{align}

\subsubsection{Achievable fidelity}

\begin{figure*}[htbp]
  \centering
  \includegraphics[width=.9\linewidth]{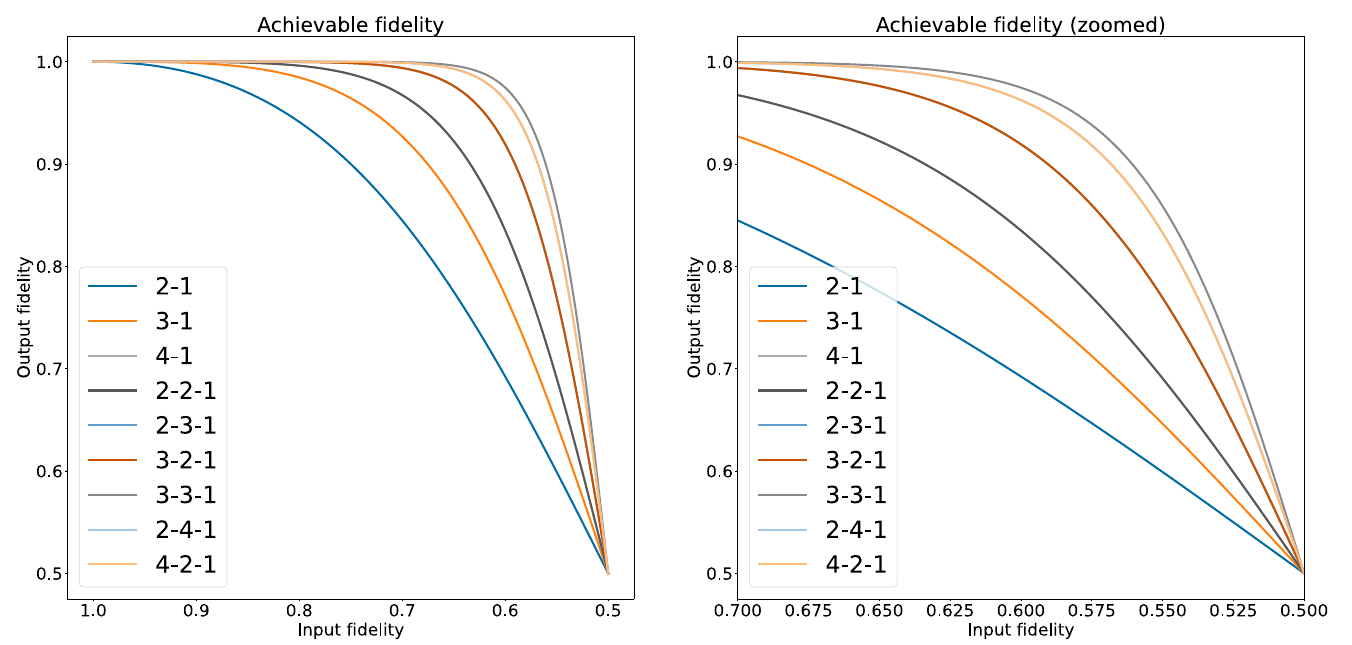}
  \caption{\textbf{Achievable fidelity - dephasing noise.} The optimal achievable fidelity for a dephased state $\rho_\text{dephase}$ for each different distillation protocol is shown. The input fidelity is $1-p_\text{dephase}$. Protocol 2-2-1 overlaps with protocol 4-1. Protocol 3-2-1 overlaps with protocol 2-3-1. Protocol 2-4-1 overlaps with protocol 4-2-1.}
  \label{fig:dephasing_fid}
\end{figure*}

Similar to the analysis of the Werner state, using more input pairs and larger ED+AD protocols produces higher output fidelity. In this case, the achievable fidelity in the low noise regime is much higher compared to the Werner state case (Fig.~6). This is due to the fact that dephasing noise is easier to correct than depolarizing noise.

\subsubsection{BB84 key rates}

\begin{figure*}[htbp]
  \centering
  \includegraphics[width=.9\linewidth]{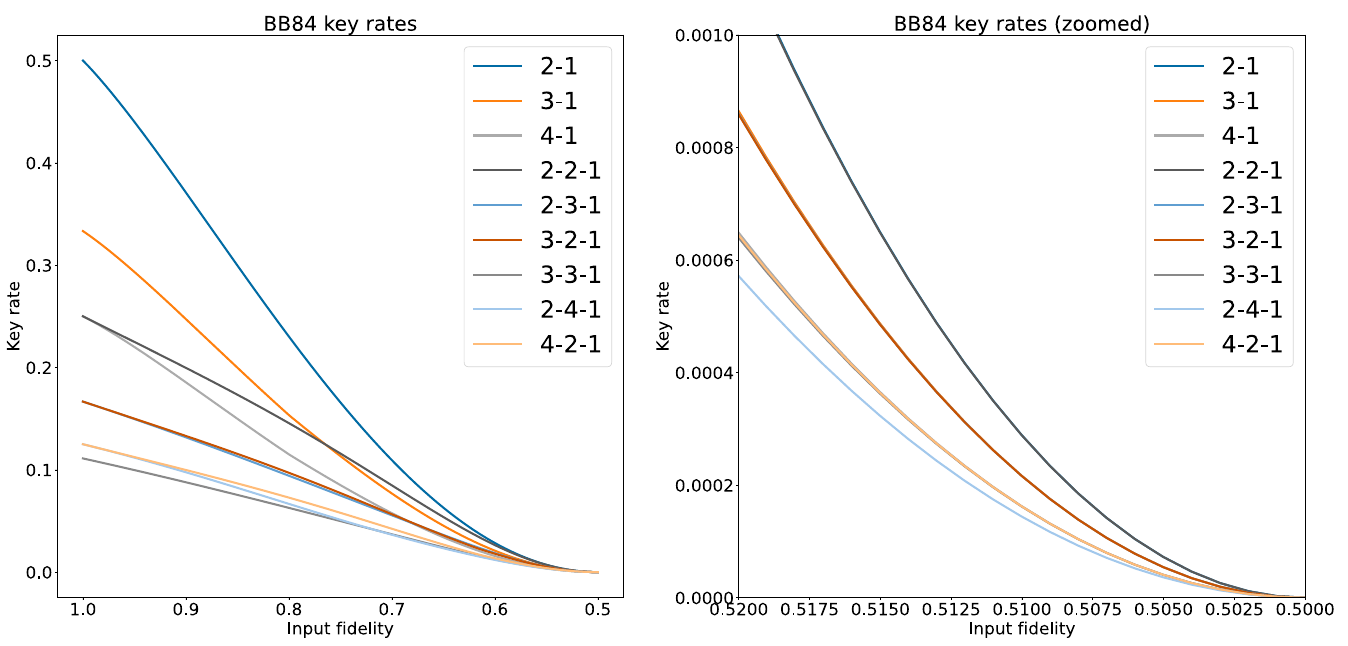}
  \caption{\textbf{Achievable BB84 asymptotic key rates - dephasing noise.} For each input fidelity, we enumerate all possible distillation protocols and plot the optimal key rate achieved by the distillation protocol. The input fidelity is $1-p_\text{dephase}$. In the zoomed figure, protocol 2-1 behaves similarly to protocol 2-2-1, protocol 3-1 behaves similarly to protocol 2-3-1 and protocol 3-2-1, and protocol 3-3-1 behaves similarly to protocol 4-1 and protocol 4-2-1. }
  \label{fig:dephasing_keyrates}
\end{figure*}
In terms of achievable key rates, the ED-only protocol $2$-$1$ outperforms all other more complicated protocols (Fig.~7) in all noise regimes. The special structure of the dephasing noise eliminates the need for complicated distillation protocols such as the $3$-$3$-$1$ protocol. While these protocols can produce high-quality entanglement (Fig.~\ref{fig:dephasing_fid}), the low success probability makes them less efficient compared to ED-only protocols. The results here demonstrate that depending on the nature of the noise in the input state as well as the strength of the noise, very different distillation protocols should be used in order to achieve optimal performance of entanglement-based QKD.

\subsection{Security bounds for ED+AD distillation protocols}
Security bounds are usually defined as the critical quantum bit error rate (QBER) so that for any larger bit error rate, key sharing becomes impossible. It is an important parameter in the security proof of QKD protocols~\cite{yuen2016security}. We first briefly review the previously derived critical QBER for two different QKD protocols. Shor and Preskill proved the unconditional security of BB84 with QBER $<0.11$ with one-way information reconciliation and privacy amplification~\cite{shor2000simple}. It was later improved to $0.124$ by Kraus, Renner, and Gisin with information-theoretical consideration~\cite{renner2005information,kraus2005lower}. On the other hand, using advantage distillation is known to improve the security bounds dramatically~\cite{gottesman2003proof}. The best security bound for BB84 with advantage distillation is $0.2$~\cite{chau2002practical, bae2007key}.

As for the six-state protocol~\cite{bruss1998optimal} (a variation of the BB84 protocol that uses three mutually unbiased bases instead of two), the best-known security bound with one-way distillation is $0.141$~\cite{kraus2005lower}, and the security bound with advantage distillation is $0.276$~\cite{chau2002practical,bae2007key}. 

Note that using entanglement distillation before the entanglement-based QKD protocol makes the security condition identical to the entanglement condition since any bi-partite entangled state can always be transformed arbitrarily close to a maximally entangled state given a large enough ED protocol. However, this might be impractical in practice, since entanglement distillation is considered costly.

Here, we derive the security bounds given a small-size ED protocol (for simplicity, we only consider using the DEJMPS protocol) and advantage distillation. In the following analysis, we closely follow the security proof used by Bae and Acin~\cite{bae2007key}. We first define the Bell-diagonal states $\rho_\text{Bell}$ with components $p_I, p_X, p_Y, p_Z$ as follow.
\begin{align}\label{eqn:bell_diagonal_states_components}
\rho_\text{Bell} &= p_I|\Phi^+\rangle \langle \Phi^+| \nonumber \\
&+ p_X|\Psi^+\rangle \langle \Psi^+| \nonumber \\
&+ p_Y|\Psi^-\rangle \langle \Psi^-| \nonumber \\
&+ p_Z|\Phi^-\rangle \langle \Phi^-|.
\end{align}

In order to share keys, the information-theoretic quantity $I(A:B)-I(A:E)$ must be positive (known as the Csiszár–Körner bound~\cite{csiszar1978broadcast}), where $I(A:B)$ is the mutual information of the probability distribution shared by Alice and Bob after a measurement and $I(A:E)$ is the mutual information between Alice and the eavesdropper, Eve. In Ref.~\cite{bae2007key}, the sufficient and necessary condition for sharing keys is derived with the state $\rho_\text{Bell}$ shared by Alice and Bob. 
Given a large but finite round of advantage distillation, the condition is
\begin{align}\label{eqn:key_condition_AD}
    \left( p_I + p_Z \right) \left( p_X + p_Y \right) < \left( p_I - p_Z \right)^2.
\end{align}

Now, we discuss the security bounds for two QKD protocols in detail separately.

\subsubsection{The six-state protocol}
We first discuss the critical QBER for the six-state protocol as it is easier to analyze due to its symmetric nature. Similar to the assumption used in Ref.~\cite{bae2007key}, given an observed QBER $Q$, Alice and Bob assume the entangled state $\rho_{AB}$ shared by them is the Werner state with the following components,
\begin{align}\label{eqn:components_six_state}
    &p_I = 1-\frac{3}{2}Q, \nonumber \\
    &p_X = \frac{Q}{2}, \nonumber \\
    &p_Y = \frac{Q}{2}, \nonumber \\
    &p_Z = \frac{Q}{2}.
\end{align}

Using the state above and evaluating the condition in Eqn.~\ref{eqn:key_condition_AD} leads to $Q<0.276$, which is the AD-only security bound for the six-state protocol. However, if we apply the DEJMPS protocol to the state in Eqn.~\ref{eqn:components_six_state} (see appendix~\ref{sec:appendix_dejmps}), we obtain the following components,

\begin{align}\label{eqn:components_six_state_DEJMPS}
    &\tilde{p}_I = \frac{5Q^2-6Q+2}{4Q^2-4Q+2}, \nonumber \\
    &\tilde{p}_X = \frac{Q^2}{4Q^2-4Q+2}, \nonumber \\
    &\tilde{p}_Y = \frac{Q^2}{4Q^2-4Q+2}, \nonumber \\
    &\tilde{p}_Z = \frac{2Q-3Q^2}{4Q^2-4Q+2}.
\end{align}

Allowing permuting the Pauli components (which can be done with local operations only) and evaluating the condition (Eqn.~\ref{eqn:key_condition_AD}) in terms of $Q$ leads to the security bound $Q<0.30$, which outperforms the AD-only security bound.

\subsubsection{The BB84 protocol}
Unlike the case of the six-state protocol, in the standard BB84 protocol, only the $z$ and $x$ bases are measured to estimate the error and obtain the raw key. Therefore, one can only assume identical QBER in $x$-basis and $z$-basis but not in $y$-basis. Similar to the analysis done in~\cite{bae2007key}, the state Alice and Bob share is assumed to be as follows.
\begin{align}\label{eqn:components_bb84}
    &p_I = 1-2Q+x, \nonumber \\
    &p_X = Q-x, \nonumber \\
    &p_Y = x, \nonumber \\
    &p_Z = Q-x,
\end{align}
where $x \in [0,Q]$. As Ref.~\cite{bae2007key} discussed, if Alice and Bob use advantage distillation for the BB84 protocol, the optimal attack by Eve with QBER $Q$ leads to $x=0$. Evaluating the condition in Eqn.~\ref{eqn:key_condition_AD} leads to $Q<0.2$, which is the AD-only security bound. 

However, if Alice and Bob are allowed to perform coherent operations on both sides, they could permute the above components to obtain:
\begin{align}\label{eqn:components_bb84_permute}
    &\tilde{p}_I = 1-2Q+x, \nonumber \\
    &\tilde{p}_X = Q-x, \nonumber \\
    &\tilde{p}_Y = Q-x, \nonumber \\
    &\tilde{p}_Z = x.
\end{align}
Evaluating the security condition Eqn.~\ref{eqn:key_condition_AD} leads to $Q<0.25$, which is also the entanglement limit QBER for the BB84 protocol~\cite{bae2007key}.

We note that the quantum operations performed by Alice and Bob in the first stage simply permute the coefficients without explicitly performing an entanglement distillation protocol. However, the critical QBER after the permutation of the coefficients is improved to the entanglement limit, which overcomes the AD-only limit ($Q=0.2$). This can be explained by the fact that, in the case of asymmetric noise, the action of permuting Pauli components before the advantage distillation can be viewed as an entanglement distillation protocol~\cite{murta2020key}.

\subsubsection{Mutual information difference with finite rounds of ED and AD}

In the previous discussion, we provided the security bounds (as critical QBER) for the six-state protocol and the BB84 protocol using our ED+AD scheme. However, it only tells us the critical QBER such that secret-key generation is possible with a large but finite number of advantage distillation rounds. In practice, Alice and Bob would refrain from using many rounds of advantage distillation as they decrease the overall key rates significantly. Therefore, we study the potentially achievable key rates with finite rounds of ED and AD. From this information, we can learn the minimal size of the distillation protocols in order to reach a positive key rate under a given noise level.

Specifically, we calculate the mutual information difference $I(A:B)-I(A:E)$ with finite rounds of entanglement distillation and advantage distillation. This has been derived previously in Ref.~\cite{bae2007key}.
\begin{align}
    I(A:B)-I(A:E) = &1- H(\epsilon_N) - (1-\epsilon_N)H\left( \frac{1-\Lambda^N_\text{eq}}{2}  \right) \nonumber \\ 
    &- \epsilon_N H\left( \frac{1-\Lambda^N_\text{diff}}{2}  \right),
    \label{eqn:mutual_info_diff}
\end{align}
where $N$ is the number of AD rounds, and
\begin{align}
    \epsilon_N &= \frac{\epsilon^N_{AB}}{\epsilon^N_{AB}+(1-\epsilon_{AB})^N}, \nonumber \\
    \epsilon_{AB} &= p_X+p_Y, \nonumber \\
    \Lambda_\text{eq} &= \frac{p_I - p_Z}{p_I + p_Z}, \nonumber \\
    \Lambda_\text{diff} &= \frac{\lvert p_X - p_Y \rvert}{p_X + p_Y}.
\end{align}

\begin{figure*}[htbp]
  \centering
  \includegraphics[width=.9\linewidth]{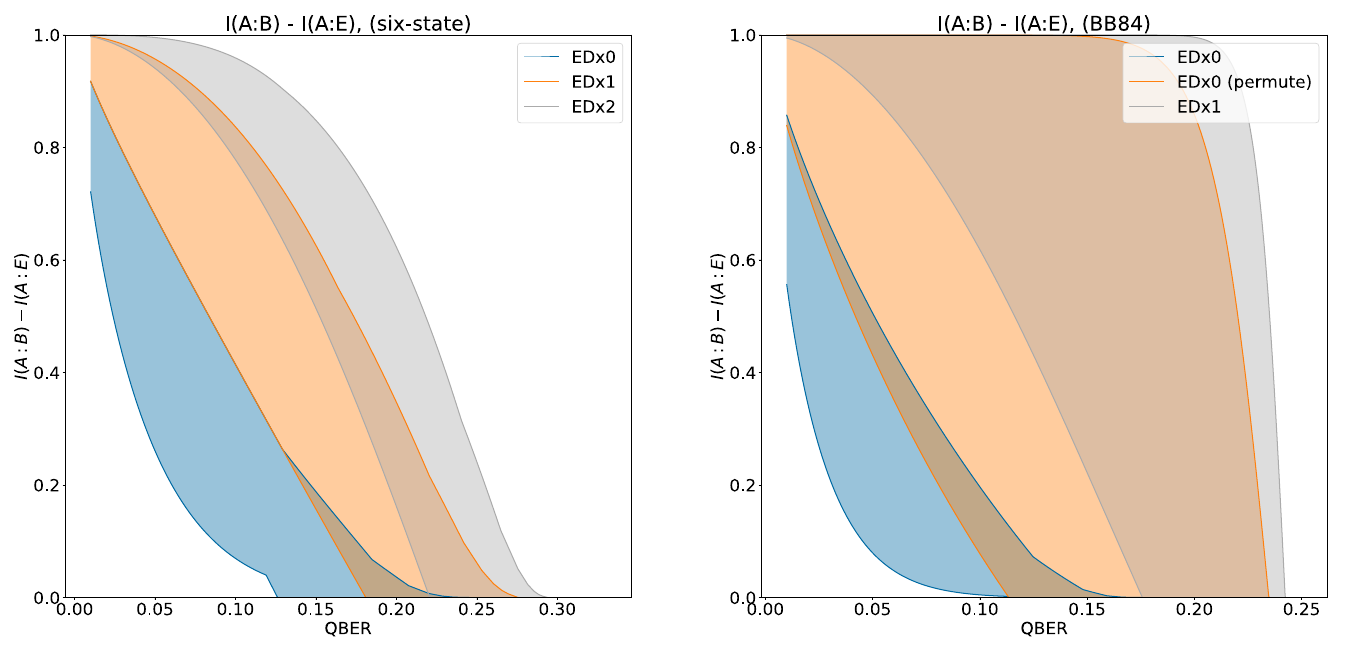}
  \caption{\textbf{Achievable key rates with finite-size ED+AD distillation.} (left) For the six-state protocol, we consider the cases of not using entanglement distillation (EDx0), using the DEJMPS protocol once (EDx1) and using the DEJMPS protocol twice (EDx2). (right) For the BB84 protocol, we consider the cases of not using entanglement distillation (EDx0), permuting the Pauli coefficients (EDx0 (permute)), and using the DEJMPS protocol once (EDx1). For both protocols, we calculate the achievable mutual information difference within 10 rounds of advantage distillation and plot as the shaded area. }
  \label{fig:finite_ED_AD}
\end{figure*}

We summarize the results in Fig.~\ref{fig:finite_ED_AD}. For both BB84 and the six-state protocol, we calculate the mutual information difference (Eqn.~\ref{eqn:mutual_info_diff}) for using different finite combinations of ED and AD. For each choice of entanglement distillation protocol, we plot the maximum and minimum of the quantities $I(A:B)-I(A:E)$ considering up to $10$ rounds of advantage distillation (solid lines). The shaded area between the solid lines is the achievable mutual information difference with a specific ED protocol given a fixed QBER.

For both BB84 and the six-state protocol, we consider three cases of using entanglement distillation (described in the caption). The non-overlapping area with increasing number of entanglement distillation indicates the advantage of using more a powerful entanglement distillation protocol with the advantage distillation. Most importantly, the results imply using a modest number of distillation resources (EDx2 for six-state protocol and permuting the coefficients for BB84 with AD up to 10 rounds) can already lead to critical QBER greater than the previously derived critical QBER ($0.2$ for BB84 and $0.276$ for the six-state protocol). In practice, Alice and Bob could estimate the QBER in the parameter estimation scheme of the protocol. Based on the QBER information, they decide the optimal ED+AD protocols in order to distill the secret keys.

The results here demonstrate the interplay between entanglement distillation, advantage distillation, and the key rate for a specific QKD protocol with our ED+AD distillation scheme and provide an estimation of the resources required for the key distillation.

\section{Conclusion}
We presented a scheme for improving entanglement-based quantum key distribution based on entanglement distillation and advantage distillation. We find both increased secret-key rates and increased resilience against noise while not imposing any additional requirements on the quantum resources. In particular, our protocol has the potential to achieve a non-zero key rate even at a noise regime where using entanglement distillation only leads to a vanishing key rate for the case of depolarizing noise.

We derived security bounds for the ED+AD distillation scheme which outperforms the previous security bounds using AD-only key distillation protocols. Moreover, the improved tolerance to the noise can be achieved within modest-size distillation protocols.

Our approach also works for more general Pauli noise models --- if one possesses knowledge about the initial noisy state shared by two parties, our framework can be immediately utilized to find the optimal ED+AD protocol for performing entanglement-based QKD. For example, applying our tools to the case of pure-dephasing noise, we found that the basic $2$-$1$ protocol achieves better key rates than more sophisticated protocols, further highlighting the practical implications of our results on near-term quantum communication.

Our work suggests several possible future directions of study. First, due to the still rather large number of all possible ED+AD protocols, studying the $m$-$n$-$1$ protocol for large $m$ and $n$ is still formidable. A discrete optimization algorithm, rather than enumerating all possible protocols, should be employed to study larger protocols.
Second, we have not considered the case of more general $m$-$n$-$o$ ED+AD protocols ($n>o>1$). In this more general case, one would need to consider all possible $n$-$o$ advantage distillation protocols to obtain optimal protocols.
Finally, in practice, one has to deal with non-identical input states and the effect of imperfect quantum operations during entanglement distillation~\cite{goodenough2024near}. We believe the framework built here can be extended to take the above considerations in mind, which in turn will yield more efficient entanglement-based QKD protocols and applications in the near future.

\appendix

\section{Binary picture and the symplectic representation of Clifford operators}\label{sec:appendix_binary_picture}
For enumerating the bi-local Clifford protocols, it will be convenient to define an equivalence relation on such protocols; equivalent protocols have equivalent distillation statistics, meaning the success probability and output state (up to local rotations) are the same. Here, we review the required mathematical framework. The material here follows Ref.~\cite{jansen2022enumerating} closely.

In the study of Clifford distillation protocols, the following binary vector representation for the Pauli group will be useful. The convention here follows Ref.~\cite{dehaene2003clifford} and Ref.~\cite{jansen2022enumerating}. We define the following mapping between binary vectors and single-qubit Pauli operators as follows
\begin{align}
  \tau_{00} \coloneqq I, \tau_{10} \coloneqq X, \tau_{11} \coloneqq Y, \tau_{01} \coloneqq Z.
\end{align}

Neglecting the scalar factors $\pm 1$, $\pm i$, we can then generalize the mapping between a Pauli string (tensor product of Pauli operators) and a binary vector to be the following
\begin{align}
  \tau_a \coloneqq \tau_{v_1 w_1} \otimes \hdots \otimes \tau_{v_n w_n}, a = \begin{bmatrix} v \\ w 
  \end{bmatrix}, v,w \in \mathbb{Z}^n_2\ .
\end{align}
Furthermore, the four Bell states can be mapped to a Pauli operator by recognizing the following relation,
\begin{align}\label{eqn:bell_states}
 |\Phi^+\rangle &= \frac{1}{\sqrt{2}} \left(|00\rangle + |11\rangle \right) = I |\Phi^+\rangle\ , \nonumber \\
 |\Psi^+\rangle &= \frac{1}{\sqrt{2}} \left(|01\rangle + |10\rangle \right) = X |\Phi^+\rangle\ , \nonumber \\
 |\Psi^-\rangle &= \frac{1}{\sqrt{2}} \left(|10\rangle - |01\rangle \right) = -iY |\Phi^+\rangle\ , \nonumber \\
 |\Phi^-\rangle &= \frac{1}{\sqrt{2}} \left(|00\rangle - |11\rangle \right) = Z |\Phi^+\rangle\ . 
\end{align}

Therefore, we can represent a tensor product of Bell states (up to a phase factor) by a binary vector. For example, the state $|\Phi^+\rangle \otimes |\Psi^+\rangle \otimes |\Psi^-\rangle \otimes |\Phi^-\rangle$ corresponds to a four-letter Pauli string $I \otimes X \otimes Y \otimes Z$, which can be mapped to a vector $a =\begin{bmatrix} 0 & 1 & 1 & 0 & 0 & 0 & 1 & 1 \end{bmatrix}^T$.

Next, a Clifford operation on $n$ qubits $C_n$ induces an automorphism on the $n$-qubit Pauli group $\mathcal{P}_n$ by conjugation $\sigma(P_n) : \mathcal{P}_n \rightarrow \mathcal{P}_n $ defined as $\sigma(P_n) = C_n P_n C_n^\dagger$. It can be shown that there exists a surjective homomorphism from the group of Clifford operators to the \emph{symplectic group} over $\mathbb{Z}_2$, $\text{Sp}(2n, \mathbb{Z}_2)$. The order of the symplectic group over $\mathbb{Z}_2$~\cite{artin2016geometric} is  
\begin{align}
 \left| \text{Sp}(2n, \mathbb{Z}_2) \right| = 2^{n^2} \prod_{j=1}^{n} (4^j - 1)\ .
\end{align}
Nevertheless, when one considers using Clifford gates to perform entanglement distillation, not all choices of Clifford operators lead to different distillation statistics. Two protocols have the same distillation statistics if the output fidelity $F=\langle \Phi^+ | \rho_{distilled} | \Phi^+ \rangle$ and the success probability are the same, and other components with respect to the Bell basis ($p_X = \langle \Psi^+ | \rho_\text{distilled} | \Psi^+ \rangle, p_Y =  \langle \Psi^- | \rho_\text{distilled} | \Psi^- \rangle, p_Z= \langle \Phi^- | \rho_\text{distilled} | \Phi^- \rangle$) are the same up to some permutation. In Ref.~\cite{jansen2022enumerating}, the equivalence classes (up to the same distillation statistics) of the symplectic group are characterized, which has a much smaller size than the original symplectic group. Let ${D}_n$ be an element in the symplectic group $\text{Sp}(2n, \mathbb{Z}_2)$ which, when left multiplied to any Clifford operator $C_n$ in the distillation protocol, does not change the distillation statistics. Such ${D}_n$ form a subgroup $\mathcal{D}_n$. When one wants to enumerate all possible bi-local Clifford protocols for a given input state, it suffices to enumerate one element in each coset of $\mathcal{D}_n$ in $\text{Sp}(2n, \mathbb{Z}_2)$, which has a size equal to the associated subgroup index 
\begin{align}
\left[ \text{Sp}(2n, \mathbb{Z}_2):\mathcal{D}_n \right] =\frac{1}{3} (2^n-1) \prod_{j=1}^{n}(2^j+1)\ .
\end{align}
The number of equivalence classes is much smaller than the size of the symplectic group. This reduction makes the enumeration of all bi-local Clifford protocols possible for a small number of shared entangled qubit pairs. 

\section{The DEJMPS protocol with Bell diagonal input states}\label{sec:appendix_dejmps}
Here, we explicitly show the output state of the DEJMPS entanglement distillation protocol (see Sec.~\ref{sec:clifford_protocols} given an input in the Bell diagonal form. (Eqn.~\ref{eqn:bell_diagonal_states_components})

We assume the input states are two identical Bell states. Hence, the state before the entanglement distillation is $\rho^{\otimes2}_\text{Bell}$, with components $\rho^{\otimes2}_{\text{Bell},ij} = \left(\rho_\text{Bell,i}\right) \left(\rho_\text{Bell,j}\right)$ for $i,j \in {I,X,Y,Z}$. Upon success, the components of the output state $\tilde{\rho}$ and the successful probability $p_\text{suc}$ are

\begin{align}
    p_\text{suc} &= p_{II} + p_{XX} + p_{ZX} + p_{YI} \nonumber \\
    &+ p_{YY} + p_{ZZ} +p_{XZ} + p_{IY}, \nonumber \\
    \tilde{\rho}_I &= \frac{\left( p_{II} + p_{YY} \right)}{p_\text{suc}}, \nonumber \\
    \tilde{\rho}_X &= \frac{\left( p_{XX} + p_{ZZ} \right)}{p_\text{suc}}, \nonumber \\
    \tilde{\rho}_Y &= \frac{\left( p_{ZX} + p_{XZ} \right)}{p_\text{suc}}, \nonumber \\
    \tilde{\rho}_Z &= \frac{\left( p_{YI} + p_{IY} \right)}{p_\text{suc}}.
\end{align}

Note that in order to maximize the output fidelity $\tilde{\rho}_I$, one should permute the input state's components such that $p_I > p_X \geq p_Z \geq p_Y$~\cite{dehaene2003local}.

\section{Comparison with DEJMPS-based protocols}\label{sec:appendix_comparison_DEJMPS}
\begin{figure}[htbp]
  \centering
  \includegraphics[width=.85\linewidth]{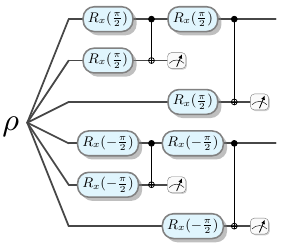}
  \caption{\textbf{Illustration of a $3$-$1$ DEJMPS protocol.}  The $3$-$1$ DEJMPS protocol is the recursive version of the standard DEJMPS protocol. The first unmeasured pair is the output of the protocol.}
  \label{fig:dejmps31}
\end{figure}

In the main text, we performed enumeration over all possible ED protocols to find the optimal ED+AD protocol for given noisy input states. Here, we demonstrate the necessity of such enumeration and show the improvement of key rate over naively using the DEJMPS protocols in the ED part. The standard $2$-$1$ DEJMPS protocol is defined in Sec.~\ref{sec:clifford_protocols}. The $m$-$1$ DEJMPS protocol refers to recursively performing $2$-$1$ DEJMPS protocol to the noisy pair. Fig.~\ref{fig:dejmps31} shows a $3$-$1$ DEJMPS protocol. 

\begin{figure}[htbp]
  \centering
  \includegraphics[width=.9\linewidth]{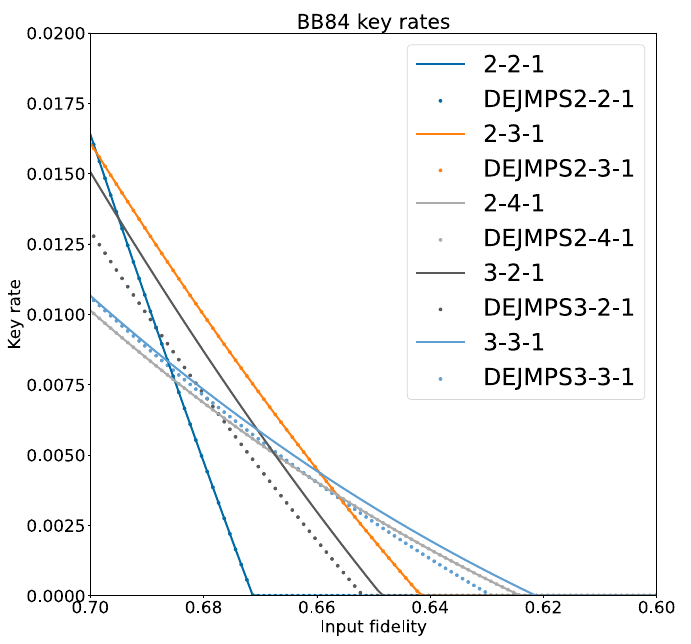}
  \caption{\textbf{Comparison of achievable BB84 key rate between optimal protocols and DEJMPS-based protocols. (Depolarizing noise)} The achievable BB84 key rates with optimal ED+AD protocols (line) and DEJMPS-based ED+AD protocols (dots) are plotted. The input state is the Werner state defined in Eqn.~\ref{eqn:werner}. }
  \label{fig:depolarizing_dejmps_keyrates}
\end{figure}

Fig.~\ref{fig:depolarizing_dejmps_keyrates} shows the achievable BB84 key rates (with Werner state input) for the optimal ED+AD protocols and their DEJMPS-based counterparts. In the case of high input fidelity, protocol $2$-$2$-$1$ has the highest key rate due to its lower overhead. However, in the heavy noise regime, larger ED+AD protocols become advantageous. Nevertheless, across all noise strengths, all optimal ED+AD protocols obtained from the enumeration perform the same or better than their DEJMPS-based counterparts. This demonstrates the advantage gained from the enumeration.

\section{Repetition code based distillation on Werner states}\label{sec:appendix_repetition_code}
Here, we provide analytical results for performing repetition code based advantage distillation on Werner state input (parameterized as in Eqn.~\ref{eqn:werner}). The result here explains the subtlety that the achievable fidelity by protocol $2$-$3$-$1$ is higher than protocol $2$-$4$-$1$ as discussed in Sec.~\ref{sec:depolarizing_fidelity}. We define $p_I$, $p_X$, $p_Y$, and $p_Z$ to be the Pauli components of a quantum state with respect to $|\Phi^+\rangle$, $|\Psi^+\rangle$, $|\Psi^-\rangle$, and $|\Phi^-\rangle$, respectively. 

For $2$-$1$ repetition code advantage distillation, the components after distillation are
\begin{align}
 p_I &= \frac{F^{2} + \left(\frac{1}{3} - \frac{F}{3}\right)^{2}}{F^{2} + 2 F \left(\frac{1}{3} - \frac{F}{3}\right) + 5 \left(\frac{1}{3} - \frac{F}{3}\right)^{2}}
 = \frac{10 F^{2} - 2 F + 1}{8 F^{2} - 4 F + 5}\ , \nonumber \\
 p_X &= \frac{2 \left(\frac{1}{3} - \frac{F}{3}\right)^{2}}{F^{2} + 2 F \left(\frac{1}{3} - \frac{F}{3}\right) + 5 \left(\frac{1}{3} - \frac{F}{3}\right)^{2}}
 = \frac{2 \left(F^{2} - 2 F + 1\right)}{8 F^{2} - 4 F + 5} \nonumber\ , \\
 p_Y &= p_X \ , \nonumber \\
 p_Z &= \frac{2 F \left(\frac{1}{3} - \frac{F}{3}\right)}{F^{2} + 2 F \left(\frac{1}{3} - \frac{F}{3}\right) + 5 \left(\frac{1}{3} - \frac{F}{3}\right)^{2}}
 = \frac{6 F \left(1 - F\right)}{8 F^{2} - 4 F + 5}\ .
\end{align}

For $3$-$1$ repetition code advantage distillation, the components after distillation are
\begin{widetext}
    \begin{align}
 p_I &= \frac{F^{3} + 3 F \left(\frac{1}{3} - \frac{F}{3}\right)^{2}}{F^{3} + 3 F^{2} \cdot \left(\frac{1}{3} - \frac{F}{3}\right) + 3 F \left(\frac{1}{3} - \frac{F}{3}\right)^{2} + 9 \left(\frac{1}{3} - \frac{F}{3}\right)^{3}}
  \nonumber \\
 &=F, \nonumber \\
 p_X &= \frac{4 \left(\frac{1}{3} - \frac{F}{3}\right)^{3}}{F^{3} + 3 F^{2} \cdot \left(\frac{1}{3} - \frac{F}{3}\right) + 3 F \left(\frac{1}{3} - \frac{F}{3}\right)^{2} + 9 \left(\frac{1}{3} - \frac{F}{3}\right)^{3}}
\nonumber \\ &= \frac{4 \left(- F^{3} + 3 F^{2} - 3 F + 1\right)}{9 \cdot \left(4 F^{2} - 2 F + 1\right)} \ , \nonumber \\
 p_Y &= p_X \ , \nonumber \\
 p_Z &= \frac{3 F^{2} \cdot \left(\frac{1}{3} - \frac{F}{3}\right) + \left(\frac{1}{3} - \frac{F}{3}\right)^{3}}{F^{3} + 3 F^{2} \cdot \left(\frac{1}{3} - \frac{F}{3}\right) + 3 F \left(\frac{1}{3} - \frac{F}{3}\right)^{2} + 9 \left(\frac{1}{3} - \frac{F}{3}\right)^{3}}
 \nonumber \\ &= \frac{- 28 F^{3} + 30 F^{2} - 3 F + 1}{9 \cdot \left(4 F^{2} - 2 F + 1\right)}\ .
\end{align}
\end{widetext}

Finally, for $4$-$1$ repetition code advantage distillation, the components after distillation are,
\begin{widetext}
    \begin{align}
 p_I &= \frac{F^{4} + 6 F^{2} \left(\frac{1}{3} - \frac{F}{3}\right)^{2} + \left(\frac{1}{3} - \frac{F}{3}\right)^{4}}{F^{4} + 4 F^{3} \cdot \left(\frac{1}{3} - \frac{F}{3}\right) + 6 F^{2} \left(\frac{1}{3} - \frac{F}{3}\right)^{2} + 4 F \left(\frac{1}{3} - \frac{F}{3}\right)^{3} + 17 \left(\frac{1}{3} - \frac{F}{3}\right)^{4}} \nonumber \\
      &= \frac{136 F^{4} - 112 F^{3} + 60 F^{2} - 4 F + 1}{32 F^{4} - 32 F^{3} + 120 F^{2} - 56 F + 17} \nonumber \\
 p_X &= \frac{8 \left(\frac{1}{3} - \frac{F}{3}\right)^{4}}{F^{4} + 4 F^{3} \cdot \left(\frac{1}{3} - \frac{F}{3}\right) + 6 F^{2} \left(\frac{1}{3} - \frac{F}{3}\right)^{2} + 4 F \left(\frac{1}{3} - \frac{F}{3}\right)^{3} + 17 \left(\frac{1}{3} - \frac{F}{3}\right)^{4}} \nonumber \\
      &= \frac{8 \left(F^{4} - 4 F^{3} + 6 F^{2} - 4 F + 1\right)}{32 F^{4} - 32 F^{3} + 120 F^{2} - 56 F + 17} \nonumber \\
 p_Y &= p_X \nonumber \\
 p_Z &= \frac{4 F^{3} \cdot \left(\frac{1}{3} - \frac{F}{3}\right) + 4 F \left(\frac{1}{3} - \frac{F}{3}\right)^{3}}{F^{4} + 4 F^{3} \cdot \left(\frac{1}{3} - \frac{F}{3}\right) + 6 F^{2} \left(\frac{1}{3} - \frac{F}{3}\right)^{2} + 4 F \left(\frac{1}{3} - \frac{F}{3}\right)^{3} + 17 \left(\frac{1}{3} - \frac{F}{3}\right)^{4}} \nonumber \\
      &= \frac{12 F \left(- 10 F^{3} + 12 F^{2} - 3 F + 1\right)}{32 F^{4} - 32 F^{3} + 120 F^{2} - 56 F + 17}.
\end{align}
\end{widetext}

\begin{figure*}[htbp]
  \centering
  \includegraphics[width=1\linewidth]{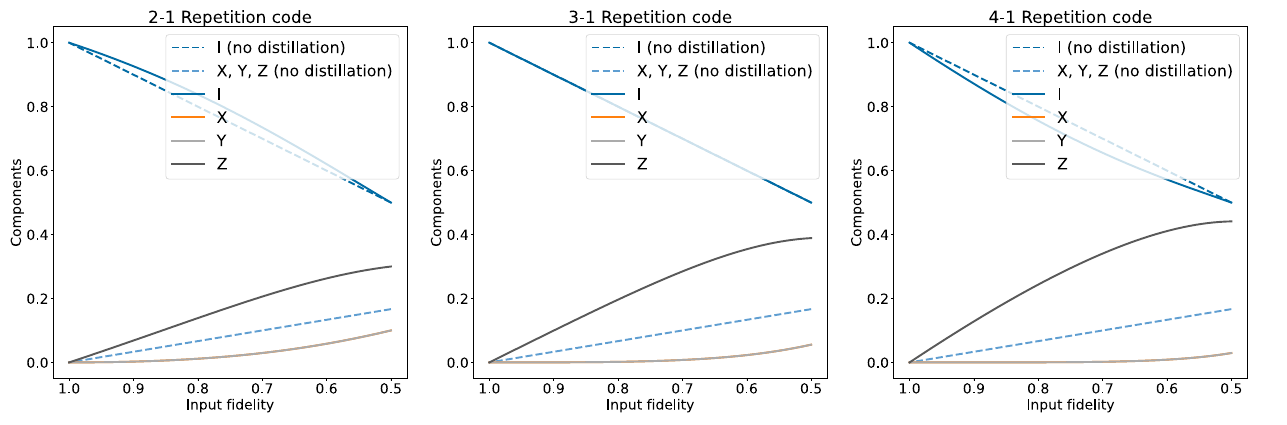}
  \caption{\textbf{Pauli components after repetition code distillation for Werner states.} The Pauli components after repetition-code-based distillation are shown. The dashed line corresponds to no distillation. Solid lines are components after distillation.}
  \label{fig:repetition_code_components}
\end{figure*}

In Fig.~\ref{fig:repetition_code_components}, we plot the components against different input fidelity. In the case of $2$-$1$ distillation, the $p_I$ (fidelity with respect to $|\Phi^+\rangle$) and $p_Z$ components are both increased. However, in the case of $3$-$1$ distillation, the $p_I$ remains before and after the distillation. For $4$-$1$ distillation, the $p_I$ component actually decreases after the distillation. Nevertheless, the bit-error components ($p_X$ and $p_Y$) are suppressed in all cases.

\begin{figure}[htbp]
  \centering
  \includegraphics[width=.9\linewidth]{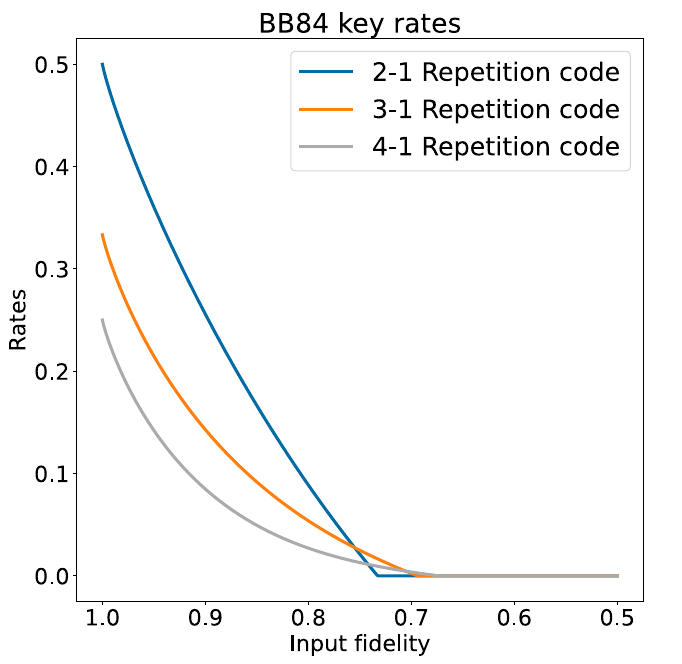}
  \caption{\textbf{Achievable BB84 key rate with repetition code distillation for Werner states.} The achievable BB84 key rates with different repetition code based distillation are shown.}
  \label{fig:repetition_code_keyrates}
\end{figure}

Although the output fidelity ($p_I$) can decrease with a larger repetition code, the BB84 key rate results (Fig.~\ref{fig:repetition_code_keyrates}) suggest that the larger code can still tolerate more noise for the actual key distribution task.

\begin{acknowledgments}
We thank Hiroki Takahashi for stimulating discussions. S. S. thanks Xiongfeng Ma, William J. Munro, and Koji Azuma for their comments regarding the security bounds.  
This project was supported by the JST Moonshot R\&D program under Grant JPMJMS226C.
\end{acknowledgments}

\bibliography{citation}

\end{document}